\documentclass[]{emulateapj}
\usepackage{apjfonts}

\usepackage{color}
\DeclareMathSymbol{\varOmega}{\mathord}{letters}{"0A}
\DeclareMathSymbol{\varSigma}{\mathord}{letters}{"06}
\DeclareMathSymbol{\varPsi}{\mathord}{letters}{"09}
\DeclareMathSymbol{\varPhi}{\mathord}{letters}{"08}

\newcommand{\Fig}[1]{Fig.~\ref{#1}}

\shorttitle{Particle Clumping and Planetesimal Formation Depend Strongly on
Metallicity}
\shortauthors{Johansen, Youdin, \& Mac Low}

\begin{document}

\title{Particle Clumping and Planetesimal Formation\\Depend Strongly on
Metallicity}

\author{Anders Johansen}
\affil{Leiden Observatory, Leiden University, P.O.\ Box 9513, 2300 RA Leiden,
The Netherlands}
\email{ajohan@strw.leidenuniv.nl}

\author{Andrew Youdin}
\affil{Canadian Institute for Theoretical Astrophysics, University of Toronto,
60 St. George Street, Toronto, Ontario M5S 3H8, Canada}

\and

\author{Mordecai-Mark Mac Low}
\affil{Department of Astrophysics, American Museum of Natural History, 79th
Street at Central Park West, New York, NY 10024-5192, USA}

\begin{abstract}
We present three-dimensional numerical simulations of particle clumping and
planetesimal formation in protoplanetary disks with varying amounts of solid
material. As centimeter-size pebbles settle to the mid-plane, turbulence
develops through vertical shearing and streaming instabilities. We find that
when the pebble-to-gas column density ratio is 0.01, corresponding roughly to
solar metallicity, clumping is weak, so the pebble density rarely exceeds the
gas density. Doubling the column density ratio leads to a dramatic increase in
clumping, with characteristic particle densities more than ten times the gas
density and maximum densities reaching several thousand times the gas density.
This is consistent with unstratified simulations of the streaming instability
that show strong clumping in particle dominated flows. The clumps readily
contract gravitationally into interacting planetesimals of order 100 km in
radius. Our results suggest that the correlation between host star metallicity
and exoplanets may reflect the early stages of planet formation. We further
speculate that initially low metallicity disks can be particle enriched during
the gas dispersal phase, leading to a late burst of planetesimal formation.
\end{abstract}


\section{Introduction}

The concentration of particles to high spatial densities promotes the formation
of planetesimals, the super-kilometer scale building blocks of planets. Drag
forces on pebbles and rocks in disks lead to spontaneous particle clumping
\citep{GoodmanPindor2000}. The discovery of a linear streaming instability
\citep{YoudinGoodman2005} shows that clumping is a robust consequence of
particles drifting in and gas flowing out in disks with some radial pressure
support \citep{Nakagawa+etal1986}. \citet{JohansenYoudin2007} studied the
non-linear saturation of the streaming instability, neglecting vertical gravity
and self-gravity. Those simulations showed that groups of boulders accelerate
the gas around them towards the Keplerian velocity, reducing the radial drift
locally and leading to temporary concentrations of boulders \citep[see
also][]{Johansen+etal2006}.
\begin{figure*}
  \begin{center}
    \includegraphics[width=0.8\linewidth]{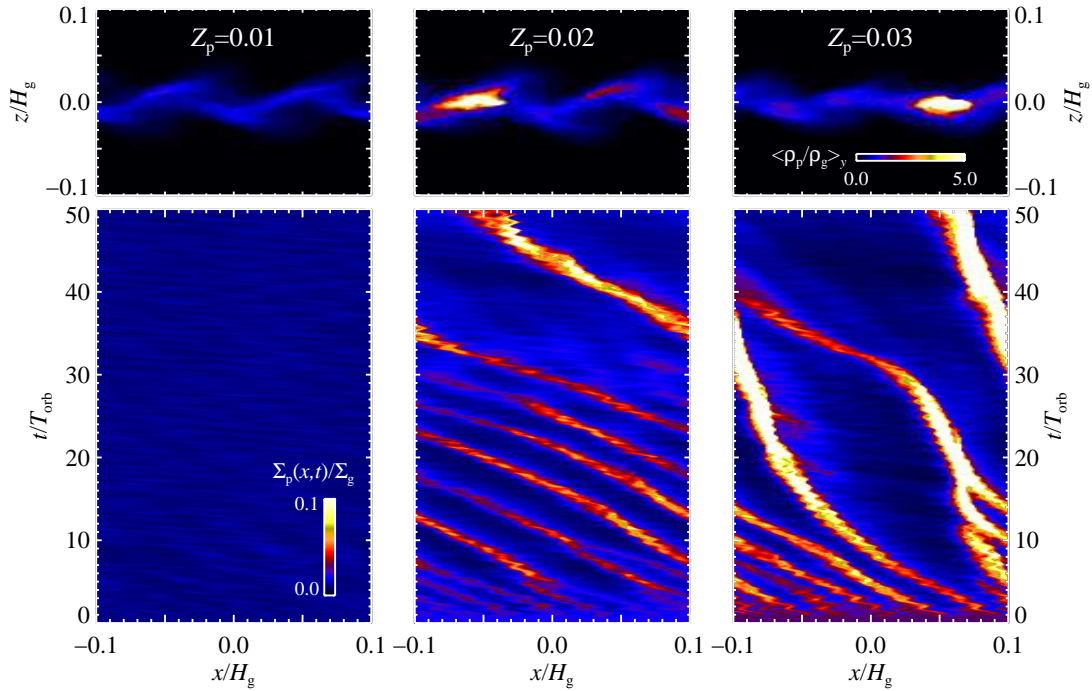}
  \end{center}
  \caption{Results of $128^3$ sedimentation simulations for various values of
  the pebble abundance $Z_{\rm p}$ (different columns). The upper row of images
  shows the particle density averaged over the azimuthal direction at the end
  of each simulation. The wave-like pattern in the particle density is a
  consequence of coherent radial drift and vertical oscillations. The lower row
  shows particle density averaged over both $y$ and $z$ directions, as a
  function of radial coordinate $x$ and time $t$. At roughly solar metallicity
  (left column) the particle surface density is uniform. Super-solar
  metallicities (middle and right columns) produce significant clumps that
  merge over time because weaker clumps drift faster.}
  \label{f:Sigmapmx_t}
\end{figure*}

\cite{Chiang2008} and \cite{Barranco2009} recently performed three-dimensional
(3D) simulations of vertical shear instabilities in Keplerian disks in the
single fluid limit where particles and gas have exactly the same velocities.
These studies confirmed expectations that mid-plane turbulence develops when
the Richardson number ${\rm Ri}\lesssim 1$. While perfect coupling is a good
approximation for small grains, it cannot include vertical settling or in-plane
streaming motions. 

In this Letter we present 3D simulations of the motion of gas and pebbles in
sub-Keplerian disks, including vertical gravity and particle sedimentation.
Thus, we can study the combined effect of vertical shearing and streaming
instabilities, as particles self-consistently settle towards -- and are stirred
from -- the mid-plane. We exclude external sources of turbulence, including
magnetorotational instabilities \citep[which can actually promote clumping,
see][]{Johansen+etal2007,Balsara+etal2009}. Our hydrodynamical simulations
offer a first approximation to dead zones with low ionization
\citep{Sano+etal2000} where turbulent surface layers drive only weak motions in
the mid-plane \citep{FlemingStone2003,Oishi+etal2007}.

In this non-magnetized limit, we investigate the clumping of smaller particles
than considered in \cite{Johansen+etal2007}, which increases the likelihood of
coagulation up to the initial sizes. We find that clumping of pebbles in the
mid-plane layer increases sharply above a threshold mass fraction of solids
roughly consistent with solar metallicity. Thus planetesimal formation may help
explain the high probability of finding giant exoplanets around stars rich in
heavy elements
\citep{Gonzalez1997,Santos+etal2001,FischerValenti2005,JohnsonApps2009}. 

\section{Simulations}

We perform 3D hybrid simulations. They model gas on a fixed grid and solids
with superparticles, each representing a swarm of actual particles. We solve
the standard shearing sheet dynamical equations for a frame rotating at the
Keplerian frequency $\varOmega$ at a fixed orbital distance $r$ from the star.
The axes are oriented such that $x$ points radially outwards, $y$ points in the
orbital direction, while $z$ points vertically out of the disk. The gas is
subject to a global radial pressure gradient that reduces the gas orbital speed
by $\Delta v=-0.05c_{\rm s}\approx 25$\,m\,s$^{-1}$
\citep{ChiangGoldreich1997}. The sound speed $c_{\rm s}$, gas scale height
$H_{\rm g} = c_{\rm s}/\varOmega$ and mid-plane gas density $\rho_0$ are the
natural units of the simulation.

The motion of gas and particles are coupled through momentum-conserving drag
forces with particle friction time $\tau_{\rm f}$. Our dynamical equations are
identical to those of \cite{YoudinJohansen2007}, with the addition of a
vertical gravitational acceleration $g_z=-\varOmega^2 z$ affecting both gas and
particles.

The superparticles are evenly distributed in mass and number into four bins of
normalized friction time $\varOmega \tau_{\rm f}=0.1,0.2,0.3,0.4$. These
friction times are characteristic of compact solids with radius $a_{\rm
p}\approx3,6,9,12$ cm at $r=5\,{\rm AU}$ in the Minimum Mass Solar Nebula
\citep{Weidenschilling1977a,Weidenschilling1977b,Hayashi1981}. Rescaling to
$r=10\,{\rm AU}$ yields $a_{\rm p} \approx 1$--4~cm. We colloquially refer to
this range of particle sizes as pebbles to contrast with larger $\varOmega
\tau_{\rm f} \approx 1.0$ boulders. The total pebble mass is fixed by setting
the pebble-to-gas column density ratio $Z_{\rm p}=\langle\varSigma_{\rm
p}\rangle/\langle\varSigma_{\rm g}\rangle$, where $\langle\varSigma_{\rm
p}\rangle$ and $\langle\varSigma_{\rm g}\rangle$ are the mean particle and gas
column densities, taking into account that most of the gas resides beyond the
vertical extent of the simulation box. This pebble abundance turns out to be
the crucial parameter for triggering particle clumping.
\begin{figure*}
  \begin{center}
    \includegraphics[width=0.8\linewidth]{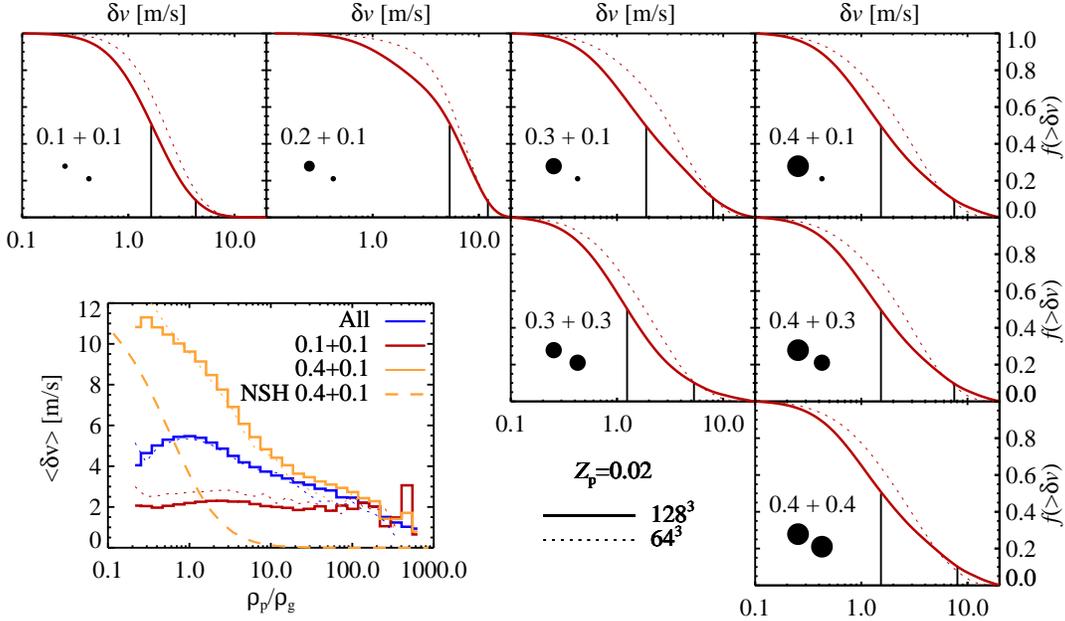}
  \end{center}
  \caption{Cumulative distribution functions of the relative speed between
  particles of various sizes, weighted by the relative speed itself. Vertical
  lines mark 50\% and 10\% in the distribution functions. The typical collision
  speed lies between 0.5 and 5\,m\,s$^{-1}$, although collisions at up to
  10\,m\,s$^{-1}$ occur as well. Collision combinations 0.2+0.2, 0.2+0.3,
  0.2+0.4 showed very similar results and are not included in the figure. The
  low resolution simulation (dotted lines) shows a bit higher collision speeds,
  because less clumping occurs in it. The lower left panel shows the mean
  collision speed as a function of the density of particles in the grid cell.
  Grid cells with a high particle density have typical collision speeds below
  2\,m\,s$^{-1}$. At low particle densities the collision speed between
  $\varOmega\tau_{\rm f}=0.1$ and $\varOmega\tau_{\rm f}=0.4$ particles rises,
  in agreement with differential drift \citep[NSH analytical
  solution,][]{Nakagawa+etal1986}, while the overall collision speed tends
  towards the collision speed between the $\varOmega\tau_{\rm f}=0.1$ bodies
  that dominate low density regions.}
  \label{f:vcoll}
\end{figure*}

The total abundance of condensable materials beyond the ice line was estimated
by \citet{Hayashi1981} to be $Z_{\rm c} \approx 0.018$, while more up-to-date
models give a somewhat lower value of $Z_{\rm c} \approx 0.015$ at temperatures
less than 41 K \citep{Lodders2003,AllendePrieto+etal2001}. For our models a
greater uncertainty is the efficiency of conversion from dust grains to
pebbles. Assuming that a majority ($\approx 2/3$) of the condensable solids are
bound in pebbles, $Z_{\rm p}=0.01$ corresponds to solar metallicity. We also
experiment with higher values of $Z_{\rm p}=0.02$ and $Z_{\rm p}=0.03$, which
are motived both by stars with super-solar metallicities and by mechanisms that
enrich the solids-to-gas ratio in disks (see \S\ref{sec:conc}). A given pebble
abundance would correspond to higher values of the metallicity if pebbles make
up a smaller fraction of the condensable material.

We use a box size of $L_x=L_y=L_z=0.2H_{\rm g}$ and resolutions of $64^3$ zones
with 125,000 particles, and $128^3$ zones with 1,000,000 particles. This
relatively small box size is chosen to capture typical wavelengths of streaming
and Kelvin-Helmholtz instabilities. The gas density is in vertical hydrostatic
equilibrium. Particle positions are initialized to give a Gaussian density
distribution around the mid-plane with scale height $H_{\rm p}=0.02 H_{\rm g}$,
while gas and particle velocities are initially set to match the drag force
equilibrium solution of \cite{Nakagawa+etal1986}.

\section{Particle Clumping}

Since the disk is initially laminar, particles settle to the disk mid-plane.
As particles collect in the mid-plane, they accelerate gas there towards the
Keplerian orbital speed. This generates vertical shear that can drive
Kelvin-Helmholtz instabilities. The velocity difference between gas and solids
also triggers streaming instabilities. The resulting turbulence halts particle
sedimentation and can lead to clumping.

In \Fig{f:Sigmapmx_t} we show results for pebble abundances $Z_{\rm p}=0.01$
(roughly solar), $Z_{\rm p}=0.02$ (super-solar) and $Z_{\rm p}=0.03$ (strongly
super-solar). The $Z_{\rm p}=0.01$ simulation shows minimal clumping, but
reveals instead a surprising interaction between settling and stirring. The
particle layer is not centered uniformly on the mid-plane, as usually assumed.
Instead the solids organize in a wave-like pattern (top left panel of
\Fig{f:Sigmapmx_t}) produced by coherent vertical oscillations combined with
radial drift. We note that the streaming instability is known to drive coherent
vertical motions in clumps of boulders \citep{JohansenYoudin2007}.  The
characteristic particle density $\tilde{\rho}_{\rm p} \equiv \langle \rho_{\rm
p}^2\rangle/\langle \rho_{\rm p}\rangle \approx 0.9 \rho_{\rm g}$ is high
enough to exert feedback on the gas. However the average mid-plane density
$\rho_{\rm p}^{\rm (mid)}=0.6 \rho_{\rm g}$ is diluted by the voids in the
wave-like pattern, and is too small to trigger strong radial clumping.

Increasing the pebble abundance to $Z_{\rm p}=0.02$ and $Z_{\rm p}=0.03$
triggers strong overdensities in the particle layer (see \Fig{f:Sigmapmx_t}).
Initially, streaming instabilities produce many azimuthally extended clumps in
the simulation box. Since the denser clumps drift inward more slowly, mergers
result in a single dominant clump. The characteristic particle density
increases due to the strong clumping, with $\tilde{\rho}_{\rm p}/\rho_{\rm g}
\approx0.9,18,73$ for $Z_{\rm p}=0.01,0.02,0.03$, respectively. The maximum
density in the box increases from less than ten to more than 2,000 times the
gas density when the pebble abundance is increased from $Z_{\rm p}=0.01$ to
$0.02$.

So why does the clumping increase so sharply for order unity changes to the
pebble abundance $Z_{\rm p}$? \Fig{f:Sigmapmx_t} (top panels) shows that the
vertical extent of the wave-shaped particle layer decreases with $Z_{\rm p}$,
because less kinetic energy is released in particle-dominated flows
\citep{JohansenYoudin2007}. Consequently the average mid-plane particle density
increases, with $\rho_{\rm p}^{\rm (mid)}/\rho_{\rm g}=0.6,2.0,9.0$ for $Z_{\rm
p}=0.01,0.02,0.03$. Strong clumping is thus expected for the higher metallicity
cases, because the streaming instability is more powerful---for linear growth
and nonlinear clumping---when the average $\rho_{\rm p}/\rho_{\rm g} \geq 1$
\citep{YoudinGoodman2005, JohansenYoudin2007}. A runaway develops as clumping
limits stirring, giving higher average densities and in turn more clumping.

We can estimate the transitional value $Z_{\rm p}^*$ above which particles
strongly clump. Assume that the sub-layer thickness is set by vertical shear
turbulence to be $H_{\rm p} \sim \Delta v/(2\varOmega)$ as appropriate for
small particles \citep{Sekiya1998,YoudinShu2002}. Then the clumping threshold
$\rho_{\rm p}^{\rm (mid)} \gtrsim \rho_{\rm g}$ requires \begin{equation}
Z_{\rm p} \gtrsim Z_{\rm p}^* = \frac{H_{\rm p}}{H_{\rm g}} = \frac{\Delta v}{2
c_{\rm s}}\, . \end{equation} For $\Delta v=0.05 c_{\rm s}$ this yields a
critical pebble abundance of $Z_{\rm p}^*\sim0.025$. The slight overestimate of
$Z_{\rm p}^*$ may arise because our pebbles settle more effectively than small
grains. The result that clumping happens for lower $Z_{\rm p}$ when $\Delta
v/c_{\rm s}$ is smaller (weaker radial pressure support) is seen in the
simulations of \cite{Johansen+etal2007}. 

The same metallicity threshold arises in particle layers with constant
Richardson number \citep{Sekiya1998}. Above $Z_{\rm p}^*$ a high density cusp
forms in the mid-plane because the shear turbulence cannot lift solids with
mass exceeding the local gas mass \citep{YoudinShu2002}. Vertical self-gravity
causes the analytic cusps to diverge, while streaming instabilities are
responsible for the strong clumping in our dynamical simulations. 

\section{Collision Speeds}

The growth of solids into planetesimals depends on collision speeds, which
affect the rates of fragmentation and coagulation \citep{BlumWurm2008}. The
threshold speed for catastrophic destruction of a meter-scale body, with the
material strength of compact rock, is around 10\,m\,s$^{-1}$ \citep{Benz2000}.
Porous aggregates, with less material strength, have an even lower threshold
for destruction. Sticking of equal-sized bodies $\gtrsim 1$ mm is not observed.
However, small grains can adhere at speeds less than $\sim 1$\,m\,s$^{-1}$
\citep{BlumWurm2000}.

From our simulation snapshots we measured relative speeds of all particle pairs
within the same grid cell, giving $(1/2)N(N-1)$ unique collisions in a grid
cell with $N$ particles. All collisions are assumed to be head on; for a random
distribution of impact parameters the normal velocity would be reduced by 50\%
compared to the relative speed. We averaged over snapshots taken between 30 and
50 orbits of the $64^3$ and $128^3$ runs with $Z_{\rm p}=0.02$.

The distribution of collision speeds between particles of various sizes is
shown in \Fig{f:vcoll}. We weight the occurrence of relative speeds by the
speed itself to reflect that high speed collisions take place more frequently
than low speed collisions. The typical collision speed lies in the interval
0.5--5\,m\,s$^{-1}$ for a sound speed $c_{\rm s}\approx500$\,m\,s$^{-1}$.
\Fig{f:vcoll} also shows that the collision speed is systematically lower in
the denser particle clumps where the typical speed falls to less than
2\,m\,s$^{-1}$. Averaging collision speeds without weighting by the speed
itself leads to much lower typical speeds, around 0.2--2\,m\,s$^{-1}$.

The low collision speeds are surprising given the sub-Keplerian speed of
25\,m\,s$^{-1}$. The differential radial drift alone is 12.7\,m\,s$^{-1}$
between $\varOmega\tau_{\rm f}=0.1$ and $\varOmega\tau_{\rm f}=0.4$ test
particles. However, the large bodies reside in dense clumps in the mid-plane
layer where the differential radial drift is much reduced and the turbulence is
weakened by mass loading (see insert in \Fig{f:vcoll}).

In simulations of particles in magnetorotational turbulence, collision speeds
are measured to be much higher, of order $\sim$10\,m\,s$^{-1}$ \citep[and well
beyond for different-sized particles,
see][]{Johansen+etal2007,Carballido+etal2008}. The more benign collision
environment in the current simulations is more resistant to destruction and may
even allow for growth by sticking. Low random velocities in high density clumps
also favor gravitational collapse on a much shorter time-scale.
\begin{figure}
  \begin{center}
    \includegraphics[width=0.9\linewidth]{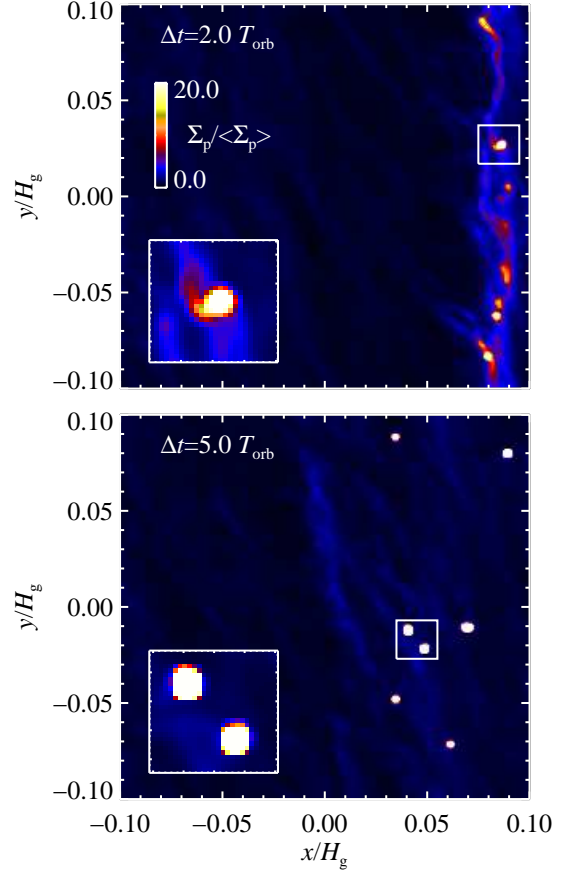}
  \end{center}
  \caption{Snapshots of particle column density after self-gravity is
  activated. A dense nearly axisymmetric filament forms by the concurrent
  action of streaming instability and self-gravity, but it continues to
  fragment into gravitationally bound clusters of pebbles and rocks. The
  clusters initially show clear signs of active accretion of solids from the
  surroundings (top insert). After 5 orbits approximately 20\% of the solid
  mass is in seven bound clusters, each containing a mass equivalent to a
  100--200-km scale solid body.}
  \label{f:planetesimals}
\end{figure}

\section{Self-gravity}

Collective gravity can act to assemble solid particles into planetesimals
\citep{Safronov1969,GoldreichWard1973,YoudinShu2002,Johansen+etal2007,Cuzzi+etal2008}.
The clumps in our $Z_{\rm p}=0.02$ and $Z_{\rm p}=0.03$ simulations reach
densities beyond the Roche density,
\begin{equation}
  \rho_{\rm R} = \frac{9}{4\pi} \frac{\varOmega^2}{G} \sim 100 \rho_{\rm g}\, ,
\end{equation}
so self-gravity will be dynamically relevant. We activate the self-gravity in a
$Z_{\rm p}=0.02$ simulation that already developed for $\sim$$40$ orbits,
starting at a chosen time that displayed relatively weak clumping. A few orbits
later a dense filament forms due to the streaming instability, but the
concurrent action of the self-gravity leads to the condensation of seven
gravitationally bound objects (see \Fig{f:planetesimals}). Initially the
planetesimals actively accrete pebbles from their surroundings. However, the
accretion slows down after 5 orbits as 20\% of the solid mass is locked in
seven clumps with masses corresponding to solid bodies of radii between 100 and
200 km. Further growth can occur both by accreting dust and pebbles and by
planetesimal-planetesimal impacts, since the modest gas density fluctuations in
mid-plane layer turbulence can not pump collision speeds of large planetesimals
into the erosive regime \citep{Ida+etal2008,Yang+etal2009}.

\section{Conclusions}\label{sec:conc}

In this Letter we present numerical evidence that the streaming instability,
which arises in the coupled motion of gas and solids, can alone provide the
necessary ingredients for the formation of planetesimals. With a high enough
abundance of pebble-sized objects, the streaming instability drives the
formation of high-density clumps in the sedimented mid-plane layer. Collision
speeds inside those clumps are relatively low, around a few meters per second.
The clumps get dense enough to contract gravitationally on a short time-scale
into gravitationally bound clusters containing mass equivalent to solid
planetesimals with radii of order 100 km.

Our simulations display a threshold dependence of clumping on the global value
of the pebble abundance. At a pebble abundance of $Z_{\rm p}=0.01$,
corresponding roughly to solar metallicity, powerful turbulence develops and
puffs up the mid-plane layer to such widths that clumping is strongly
suppressed. This environment is not conducive to planetesimal growth, with low
particle densities and high collision speeds. However, a doubling of the pebble
abundance to $Z_{\rm p}=0.02$ leads to strong clumping with local peak particle
densities exceeding a thousand times the gas density. This threshold behavior
arises because the streaming instability triggers strong clumping when the
particle density exceeds the gas density in the mid-plane.

If the fraction of condensables contained in pebbles is less than we assumed in
this discussion (approximately $2/3$), then the threshold we describe
corresponds to higher values of the bulk metallicity.
Coagulation-fragmentation models \citep{Brauer+etal2008} and observations of
pebbles in protoplanetary discs \citep{Wilner+etal2005} will be crucial for
putting tighter constraints on the pebble abundance and the location of the
metallicity switch.

The pebble abundance may increase as solid particles drift radially inward,
converging in the inner few dozen AUs of the protoplanetary disk
\citep{StepinskiValageas1996,YoudinChiang2004}. A higher $Z_{\rm p}$ also will
occur if the gas in the disk preferentially evaporates. For example,
photoevaporation removes gas from the disk surface, leaving behind larger
solids that have sedimented to the mid-plane
\citep{ThroopBally2005,AlexanderArmitage2007}. This opens a window for late
planet formation with planetesimals forming as the gas dissipates. The planets
resulting from planetesimals formed so late will not have time enough to
accrete a large gaseous envelope before the gas dissipates and would also
suffer less severe type I migration \citep{YoudinShu2002}. This could explain
why super-Earths and exo-Neptunes seem able to form equally well around stars
of low and high metallicity \citep{Udry+etal2006}.

The correlation between host star metallicity and giant exoplanets can be
partly attributed to the efficiency of forming several-Earth-mass cores from
planetesimals \citep{IdaLin2004,Mordasini+etal2009}. Our findings show that
high metallicity already facilitates planetesimal formation, giving further
support to the core accretion model for giant planet formation
\citep{Pollack+etal1996}.

\acknowledgments

Computer simulations were performed at the Huygens cluster in Amsterdam, funded
by the NWO, and at the PIA cluster of the Max Planck Institute for Astronomy.
AY and AJ are grateful to Chris Thompson for stimulating discussions at
C.I.T.A. We thank the referee for careful reading of the manuscript. M-MML was
partially supported by the NASA Origins of Solar Systems Program under grant
NNX 07-AI74G, and by the NSF Cyberenabled Discovery Initiative under grant
AST08-35734.


\begin{thebibliography}

\bibitem[Alexander \& Armitage(2007)]{AlexanderArmitage2007}
  Alexander, R.~D., \& Armitage, P.~J. 2007, \mnras, 375, 500
\bibitem[Allende Prieto et~al.(2001)Allende Prieto, Lambert, \&
Asplund]{AllendePrieto+etal2001}
  Allende Prieto, C., Lambert, D.~L., \& Asplund, M. 2001, \apjl, 556, L63
\bibitem[Balsara et~al.(2009)]{Balsara+etal2009}
  Balsara, D.~S., Tilley, D.~A., Rettig, T., \& Brittain, S.~A.,
  2009, \mnras, 397, 24
\bibitem[Barranco(2009)]{Barranco2009}
  Barranco, J.~A. 2009, \apj, 691, 907
\bibitem[Benz(2000)]{Benz2000}
  Benz, W. 2000, Space Science Reviews, 92, 279
\bibitem[Blum \& Wurm(2000)]{BlumWurm2000}
  Blum, J., \& Wurm, G. 2000, Icarus, 143, 138
\bibitem[Blum \& Wurm(2008)]{BlumWurm2008}
  Blum, J., \& Wurm, G. 2008, \araa, 46, 
\bibitem[Brauer et~al.(2008)Brauer, Dullemond, \& Henning]{Brauer+etal2008}
  Brauer, F., Dullemond, C.~P., \& Henning, T. 2008, \aap, 480, 859
\bibitem[Carballido et~al.(2008)Carballido, Stone, \& Turner]{Carballido+etal2008}
  Carballido, A., Stone, J.~M., \& Turner, N.~J. 2008, \mnras, 386, 145
\bibitem[Chiang \& Goldreich(1997)]{ChiangGoldreich1997}
  Chiang, E.~I., \& Goldreich, P. 1997, \apj, 490, 368
\bibitem[Chiang(2008)]{Chiang2008}
  Chiang, E. 2008, \apj, 675, 1549
\bibitem[Cuzzi et~al.(2008)Cuzzi, Hogan, \& Shariff]{Cuzzi+etal2008}
  Cuzzi, J.~N., Hogan, R.~C., \& Shariff, K. 2008, \apj, 687, 1432
\bibitem[Fischer \& Valenti(2005)]{FischerValenti2005}
  Fischer, D.~A., \& Valenti, J. 2005, \apj, 622, 1102
\bibitem[Fleming \& Stone (2003)]{FlemingStone2003}  Fleming, T. \&
  Stone, J. M. 2003, \apj, 585, 908
\bibitem[Goldreich \& Ward(1973)]{GoldreichWard1973}
  Goldreich, P., \& Ward, W.~R. 1973, \apj, 183, 1051
\bibitem[Gonzalez(1997)]{Gonzalez1997}
  Gonzalez, G. 1997, \mnras, 285, 403
\bibitem[Goodman \& Pindor(2000)]{GoodmanPindor2000}
  Goodman, J., \& Pindor, B. 2000, Icarus, 148, 537
\bibitem[Hayashi(1981)]{Hayashi1981}
  Hayashi, C. 1981, Prog.~Theor.~Phys., 70, 35
\bibitem[Ida \& Lin(2004)]{IdaLin2004}
  Ida, S., \& Lin, D.~N.~C. 2004, \apj, 616, 567
\bibitem[Ida et~al.(2008)Ida, Guillot, \& Morbidelli]{Ida+etal2008}
  Ida, S., Guillot, T., \& Morbidelli, A. 2008, \apj, 686, 1292
\bibitem[Johansen et~al.(2006)Johansen, Henning, \& Klahr]{Johansen+etal2006}
  Johansen, A., Henning, T., \& Klahr, H. 2006, \apj, 643, 1219
\bibitem[Johansen \& Youdin(2007)]{JohansenYoudin2007}
  Johansen, A., \& Youdin, A. 2007, \apj, 662, 627
\bibitem[Johansen et~al.(2007)Johansen, Oishi, Mac Low, Klahr, Henning, \& Youdin]{Johansen+etal2007}
  Johansen, A., Oishi, J.~S., Mac Low, M.-M., Klahr, H., Henning, T., \&
  Youdin, A. 2007, \nat, 448, 1022
\bibitem[Johnson \& Apps(2009)]{JohnsonApps2009}
  Johnson, J.~A., \& Apps, K. 2009, \apj, 699, 933
\bibitem[Lodders(2003)]{Lodders2003}
  Lodders, K. 2003, \apj, 591, 1220
\bibitem[Mordasini et~al.(2009)Mordasini, Alibert, Benz, \& Naef]{Mordasini+etal2009}
  Mordasini, C., Alibert, Y., Benz, W., \& Naef, D. 2009, \aap, 501, 1161
\bibitem[Nakagawa et~al.(1986)Nakagawa, Sekiya, \& Hayashi]{Nakagawa+etal1986}
  Nakagawa, Y., Sekiya, M., \& Hayashi, C. 1986, Icarus, 67, 375
\bibitem[Oishi et al.(2007)]{Oishi+etal2007}
  Oishi, J.~S., Mac Low, M.-M., \& Menou, K.\ 2007, \apj, 670, 805
\bibitem[Pollack et~al.(1996)]{Pollack+etal1996}
  Pollack, J.~B., Hubickyj, O., Bodenheimer, P., Lissauer, J.~J.,
  Podolak, M., \& Greenzweig, Y. 1996, Icarus, 124, 62
\bibitem[Safronov(1969)]{Safronov1969}
  Safronov, V.~S. 1969, Evoliutsiia doplanetnogo oblaka.
  (English transl.: Evolution of the Protoplanetary Cloud and Formation of
  Earth and the Planets, NASA Tech. Transl. F-677,
  Jerusalem: Israel Sci. Transl. 1972)
\bibitem[Sano et~al.(2000)Sano, Miyama, Umebayashi, \& Nakano]{Sano+etal2000}
  Sano, T., Miyama, S.~M., Umebayashi, T., \& Nakano, T. 2000, \apj, 543, 486
\bibitem[Santos et~al.(2001)Santos, Israelian, \& Mayor]{Santos+etal2001}
  Santos, N.~C., Israelian, G., \& Mayor, M. 2001, \aap, 373, 1019
\bibitem[Sekiya(1998)]{Sekiya1998}
  Sekiya, M. 1998, Icarus, 133, 298
\bibitem[Stepinski \& Valageas(1996)]{StepinskiValageas1996}
  Stepinski, T.~F., \& Valageas, P. 1996, \aap, 309, 301
\bibitem[Throop \& Bally(2005)]{ThroopBally2005}
  Throop, H.~B., \& Bally, J. 2005, \apjl, 623, L149
\bibitem[Udry et~al.(2006)]{Udry+etal2006}
  Udry, S., Mayor, M., Benz, W., Bertaux, J.-L., Bouchy, F., Lovis,
  C., Mordasini, C., Pepe, F., Queloz, D., \& Sivan, J.-P.
  2006, \aap, 447, 361
\bibitem[Weidenschilling(1977a)]{Weidenschilling1977a}
  Weidenschilling, S.~J. 1977a, \mnras, 180, 57
\bibitem[Weidenschilling(1977b)]{Weidenschilling1977b}
  Weidenschilling, S.~J. 1977b, \apss, 51, 153
\bibitem[Wilner et~al.(2005)Wilner, D'Alessio, Calvet, Claussen, \&
Hartmann]{Wilner+etal2005}
  Wilner, D.~J., D'Alessio, P., Calvet, N., Claussen, M.~J., \& Hartmann, L.
  2005, \apjl, 626, L109
\bibitem[Yang et al.(2009)]{Yang+etal2009} 
  Yang, C.-C., Mac Low, M.-M., \& Menou, K. 2009, \apj, submitted
  (arXiv:0907.1897)
\bibitem[Youdin \& Shu(2002)]{YoudinShu2002}
  Youdin, A.~N., \& Shu, F.~H. 2002, \apj, 580, 494
\bibitem[Youdin \& Chiang(2004)]{YoudinChiang2004}
  Youdin, A.~N., \& Chiang, E.~I. 2004, \apj, 601, 1109
\bibitem[Youdin \& Goodman(2005)]{YoudinGoodman2005}
  Youdin, A.~N., \& Goodman, J. 2005, \apj, 620, 459
\bibitem[Youdin \& Johansen(2007)]{YoudinJohansen2007}
  Youdin, A., \& Johansen, A. 2007, \apj, 662, 613

\end{thebibliography}
\end{document}